\documentclass[12pt,letter]{article}
\pdfoutput=1
\usepackage{graphicx, epsfig, color,cite}
\usepackage{amsmath}
\usepackage{amssymb}
\usepackage{float}
\usepackage{subfig}
\usepackage{hyperref}

\def\eslt{\not\!\!\!{E_T}}
\def\to{\rightarrow}

\def\bi{\begin{itemize}}
\def\ei{\end{itemize}}
\def\te{\tilde e}
\def\tchi{\tilde\chi}

\def\tu{\tilde u}

\def\tb{\tilde b}

\def\tst{\tilde t}
\def\ttau{\tilde \tau}

\def\tg{\tilde g}
\def\tnu{\tilde\nu}
\def\tell{\tilde\ell}
\def\tq{\tilde q}
\def\alt{\lesssim}
\def\agt{\gtrsim}
\def\be{\begin{equation}}  
\def\ee{\end{equation}}  
\def\bea{\begin{eqnarray}}  
\def\eea{\end{eqnarray}}

\begin{document}
\begin{titlepage}
\begin{flushright}
OU-HEP-251104
\end{flushright}

\vspace{0.5cm}
\begin{center}
  {\Large \bf Natural supersymmetry at a muon collider}\\
\vspace{1.2cm} \renewcommand{\thefootnote}{\fnsymbol{footnote}}
{\large Howard Baer$^{1}$\footnote[1]{Email: baer@ou.edu },
Vernon Barger$^2$\footnote[2]{Email: barger@pheno.wisc.edu},
Jessica Bolich$^1$\footnote[3]{Email: Jessica.R.Bolich-1@ou.edu},\\
Dibyashree Sengupta$^{3,4}$\footnote[4]{Email: Dibyashree.Sengupta@roma1.infn.it}and
Kairui Zhang$^1$\footnote[5]{Email: kzhang25@ou.edu}
}\\ 
\vspace{1.2cm} \renewcommand{\thefootnote}{\arabic{footnote}}
{\it 
$^1$Homer L. Dodge Department of Physics and Astronomy,
University of Oklahoma, Norman, OK 73019, USA \\[3pt]
}
{\it 
$^2$Department of Physics,
University of Wisconsin, Madison, WI 53706 USA \\[3pt]
}
{\it
$^3$ INFN, Laboratori Nazionali di Frascati,
Via E. Fermi 54, 00044 Frascati (RM), Italy} \\[3pt]
{\it
$^4$ INFN, Sezione di Roma, c/o Dipartimento di Fisica, Sapienza Università di Roma, Piazzale Aldo Moro 2, I-00185 Rome, Italy}
\end{center}

\vspace{0.5cm}
\begin{abstract}
\noindent

There is great interest within the particle physics community for
building a $\mu^+\mu^-$ collider with center-of-mass (CoM) energies ranging from
$\sqrt{s}\sim 1-14$ TeV.
For Beyond-the-Standard-Model (BSM) physics, natural supersymmetry
seems perhaps the most motivated, plausible extension of the Standard Model.
Here, we examine what can be accomplished by a muon collider with regards
to natural SUSY at various muon collider CoM energies.
In natural SUSY-- especially in the guise that would emerge from the
string landscape-- one expects sparticles to be spread over two orders of
magnitude in mass values.
A muon collider with highly variable beam energies would be most useful for
targeting 2-body reaction thresholds and Higgs boson resonances.

\end{abstract}
\end{titlepage}

\section{Introduction}
\label{sec:intro}

Perhaps the most important lesson learned from over 10 years of LHC running is
that the Standard Model (SM) reigns supreme at energy scales up to a few
hundred GeV. Nonetheless, if we follow Dirac's advice and follow where the
(SM) equations lead us, then the gauge hierarchy problem, as exemplified by
the Higgs mass quadratic divergences, still remains as a reminder that
the SM is likely incomplete.

\subsection{Status of weak scale supersymmetry}

Weak scale supersymmetry (SUSY) emerged already in the early 1980s as
perhaps an inevitable solution to the Big Hierarchy Problem in that
it is free of the destabilizing quadratic divergences that plague the
Higgs mass\cite{Witten:1981nf,Kaul:1981wp}.
The $N=1$ spacetime SUSY naturally emerges in Calabi-Yau
string compactifications on Ricci-flat manifolds of special holonomy\cite{Candelas:1985en}.
Non-SUSY compactifications have been argued to lead to instabilities
such as Witten's bubble of nothing\cite{Acharya:2019mcu}.

In the pre-LHC era, naturalness considerations seemed to imply that
superpartner masses within the context of the Minimal Supersymmetric
Standard Model (MSSM) should lie at or around the weak scale\cite{Feng:2013pwa}.
When supersymmetric matter failed to turn up in early LHC searches,
greater scrutiny was placed on the naturalness question.
\bi
\item The log-derivative measure of finetuning
$\Delta_{p_i}\equiv max_i|\partial\log m_Z^2/\partial\log p_i |$ asked for small
variation in $m_Z^2$ with respect to variation in high scale soft SUSY breaking
parameters. However, neglecting correlations between the soft parameters--
which in popular models such as mSUGRA/CMSSM are input as independent
parameters of ignorance whereas in explicit constructs they all emerge as
computable numbers in terms of the gravitino mass $m_{3/2}$-- leads to
finetuning overestimates\cite{Baer:2013gva} in the range of
$10-10^3$\cite{Baer:2023cvi}.
\item Likewise, the measure $\Delta_{HS}\equiv \delta m_{H_u}^2/m_h^2$
  neglects the fact that the logarithmic evolution of $m_{H_u}^2$ is
  dependent on the value of the soft term $m_{H_u}^2$ itself, and furthermore
  this evolution, to be phenomenologically viable, must be large enough
  to result in a (radiative-driven) breakdown of  electroweak symmetry.
  Use of $\Delta_{HS}$ also results in finetuning overestimates by
  factors of $\sim 10-10^3$\cite{Baer:2023cvi}.
\ei

A more conservative, model-independent measure of finetuning $\Delta_{EW}$
was instead invoked in Ref. \cite{Baer:2012up,Baer:2012cf}.
Minimization of the scalar potential in the MSSM allows one to relate the measured value of $m_Z$ to the MSSM Lagrangian parameters:
\be
m_Z^2/2=\frac{m_{H_d}^2+\Sigma_d^d-(m_{H_u}^2+\Sigma_u^u)\tan^2\beta}{\tan^2\beta -1}
  -\mu^2\simeq -m_{H_u}^2-\mu^2-\Sigma_u^u(\tst_{1,2})
  \label{eq:mzs}
\ee
where $m_{H_u}^2$ and $m_{H_d}^2$ are soft SUSY breaking Higgs masses,
$\mu$ is the (SUSY-conserving) Higgs/higgsino superpotential mass term\footnote{Twenty solutions to the SUSY $\mu$ problem are reviewed in Ref. \cite{Bae:2019dgg}.},
$\tan\beta \equiv v_u/v_d$ is the ratio of Higgs field vacuum expectation values
and the $\Sigma_u^u$ and $\Sigma_d^d$ terms contain an assortment of radiative
corrections, the most important of which are usually the
$\Sigma_u^u(\tst_{1,2})$\cite{Baer:2012cf}.
The numerical measure $\Delta_{EW}$ is defined by
\be
\Delta_{EW}\equiv max_i| i^{th}\ term\ on\ RHS\ of\ Eq.~\ref{eq:mzs}|/(m_Z^2/2) .
\ee
A value of $\Delta_{EW}\alt 30$ requires each independent contribution
to $m_Z$ to be within a factor of four of $m_Z$ itself.

Several implications of $\Delta_{EW}\alt 30$ are the following.
\bi
\item Since the $\mu$ parameter is directly bounded, then the higgsino-like
  electroweakinos of the MSS should have mass values $m_{\tchi_{1,2}^0}$ and
  $m_{\tchi_1^\pm}$ $\sim 100-350$ GeV. Thus, unlike most $20^{th}$ century SUSY
  constructs, the higgsinos are the lightest SUSY particles.
\item The weak scale value of $m_{H_u}^2$ is radiatively-driven to small
  negative values $\sim -(100-350\ {\rm GeV})^2$. This usually requires
  non-universal Higgs masses, as is generically expected in actual
  supergravity constructs\cite{Soni:1983rm,Kaplunovsky:1993rd,Brignole:1993dj}.
\item The (usually) dominant radiative corrections $\Sigma_u^u (\tst_{1,2})$
  are minimized by large, negative trilinear soft terms $A_0$\cite{Baer:2012up}.
  Large trilinears are also generic in SUGRA constructs.
  They also help to uplift the light Higgs mass $m_h\to 125$ GeV\cite{Carena:2002es}.
  Under $\Delta_{EW}$, top squarks may range up to $m_{\tst_1}\alt 3$ TeV
  and $m_{\tst_2}\alt 8$ TeV\cite{Baer:2015rja}.
\item The gluinos contribute to the $\Sigma_{u,d}^{u,d}$ terms at two-loop
  level and thus can range up to 6-9 TeV (model-dependent\cite{Baer:2015rja,Baer:2018hpb}).
\item First/second generation squark and slepton mass contributions
  to the weak scale are Yukawa-coupling suppressed. Thus, they can range up
  to 10-40 TeV. These latter values can lead to a mixed decoupling/quasi-degeneracy solution to the SUSY flavor and CP problems\cite{Baer:2019zfl}
  which are otherwise endemic to gravity mediation models.
  \ei

While the above considerations do not preclude superparticles at or around
the weak scale $\tilde{m}\sim 100$ GeV, LHC mass bounds and the Higgs mass
$m_h\simeq 125$ GeV require much higher values.
In addition, TeV-scale or higher values for soft SUSY breaking terms are
preferred by the landscape of vacua which emerge in modern string theory.
Restricting ourselves to hidden sector SUSY breaking via a single $F$-term,
then Douglas' argument is that nothing in string theory prefers any single
$F$-term value over any other\cite{Douglas:2004qg}.
The $F$ terms should thus be distributed as
random {\it complex} numbers in the landscape (and $D$-terms would be
distributed as random real numbers).
And since the hidden-sector SUSY breaking scale is determined as
\be
m_{hidden}^4=\sum_i F_iF_i^\dagger +\sum_\alpha D_\alpha D_\alpha
\ee
(including the possibility of additional $F_i$- and $D_\alpha$-term
breaking fields),
then the soft terms, related as $m_{soft}\sim m_{hidden}^2/m_P$, should
  be statistically distributed as a power-law
  $f_{SUSY}\sim m_{soft}^{2n_F+n_D-1}$ in the landscape\cite{Douglas:2004qg,Susskind:2004uv,Arkani-Hamed:2005zuc}.
  The textbook case
  of SUSY breaking via a single $F$ term field would thus lead to a linear
  statistical selection of large soft terms. An upper bound on soft term
  selection comes from requiring the derived value of the weak scale
  in each viable pocket-universe (PU) within the landscape to lie within the
  ABDS window\cite{Agrawal:1997gf}: $m_Z^{OU}/2\alt m_Z^{PU}\alt (2-5)m_Z^{OU}$ where
  $m_Z^{OU}=91.2$ GeV in {\it our universe}. The ABDS window coincides with
  $\Delta_{EW}\alt 30$ for pocket universes with no finetuning (random
  selection of values).
  The gist of this is that the string landscape prefers large values of
  soft terms over small values so long as the derived value of the weak scale is not too large, {\it i.e.} $\Delta_{EW}\alt 30$.
  This then coincides with the non-appearance of sparticles at LHC, and with
  the measured value of the Higgs mass\cite{Baer:2017uvn}.

\begin{figure}[htb!]
\centering
    {\includegraphics[height=0.4\textheight]{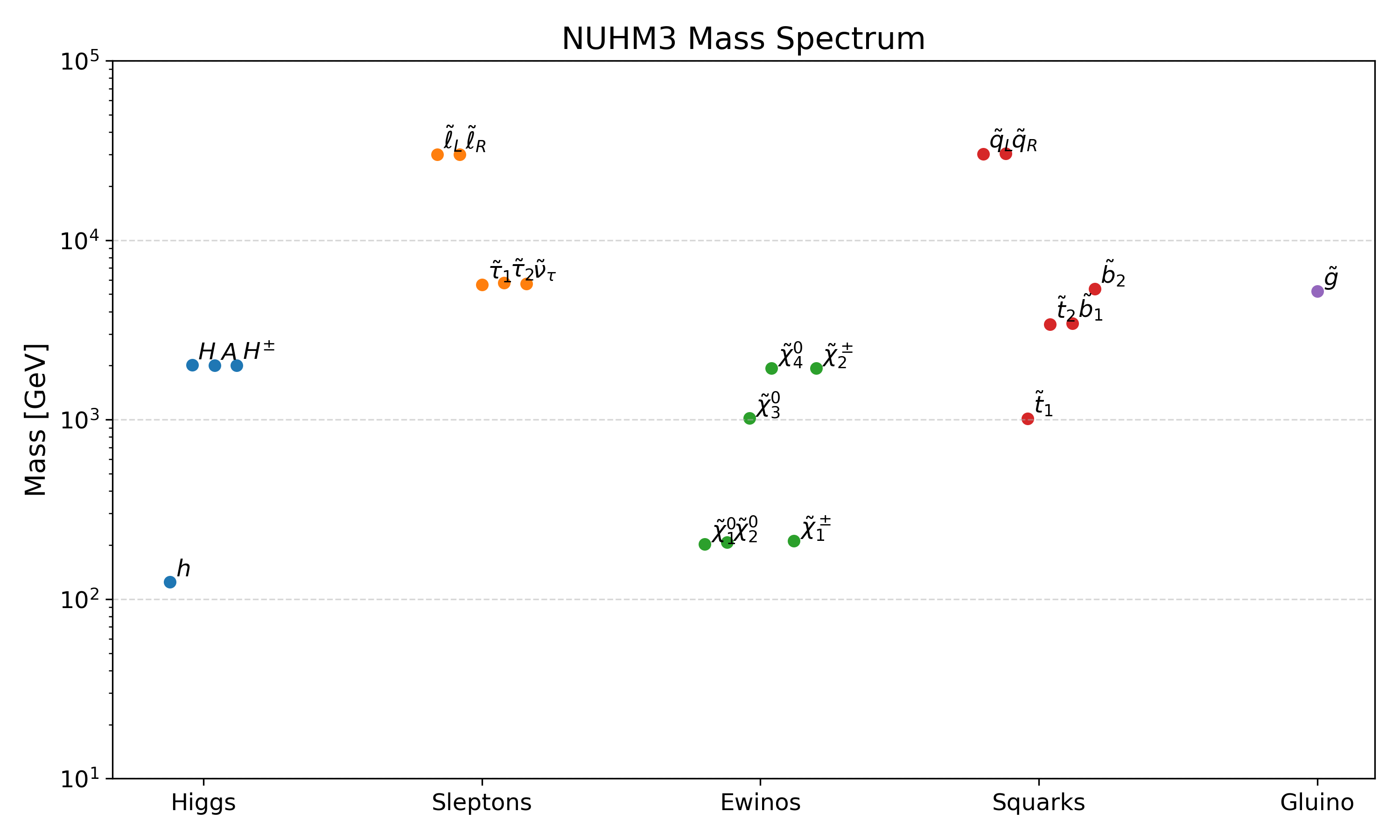}}
    \caption{Plot of typical natural SUSY mass spectra as expected from
      the string theory landscape.
      \label{fig:spec}}
\end{figure}

  The expected natural SUSY sparticle mass spectra is thus shown in
  Fig. \ref{fig:spec}, and would be typically spread over two orders of
  magnitude in values. The vector bosons $W$ and $Z$, Higgs boson $h$ and
  higgsinos should all lie in the 100-300 GeV range. Binos, winos and
  light top squarks $\tst_1$ could lie in the few-TeV range while $\tst_2$,
  $\tb_{1,2}$, gluinos and heavy Higgs $H,A$ and $H^\pm$ could lie within
  the 3-8 TeV range.
  First/second generation squarks and sleptons would likely lie in the
  10-40 TeV range.
  Thus, the superparticle mass spectrum would likely be spread out over a
  range of two orders of magnitude. Given such a situation, how does this
  information affect plans for future accelerator projects throughout the world?

\subsection{Future accelerator facilities and a $\mu^+\mu^-$ collider}

The CERN high-luminosity LHC (HL-LHC) stands out as the forefront
particle physics accelerator facility over the next decade.
Recently, a review of HL-LHC capabilities for SUSY discovery in
so-called plausible (natural) SUSY models was recently posted.
In Ref. \cite{Baer:2025zqt}, it is found that the most lucrative avenues for SUSY
discovery at HL-LHC are via higgsino pair production (taking advantage of the relatively light higgsinos with $\mu\sim 100-350$ GeV).
In particular, the soft-opposite-sign dilepton plus jet(s) plus $\eslt$
(SOSDJMET) signal
arising from $pp\to\tchi_1^0\tchi_2^0$ or $\tchi_1^\pm\tchi_2^0$
followed by $\tchi_2^0\to\tchi_1^0\ell^+\ell^-$ seems most lucrative\cite{Baer:2011ec}
(where an initial state jet radiation is used as a trigger)\cite{Han:2014kaa,Baer:2014kya,Han:2015lma}.
In fact, both ATLAS\cite{ATLAS:2019lng} and CMS\cite{CMS:2021edw} have some excess in this channel
from Run 2 data, and it will be exciting to see what Run 3 has in store.
HL-LHC reach projections in this channel are shown in Ref. \cite{Baer:2021srt}.
While HL-LHC can probe much of the natural parameter of the higgsino
discovery plane\cite{Baer:2020sgm}, some parameter space with higher $\mu\sim 250-350$ GeV
seems out of reach.

Another lucrative search channel for HL-LHC is via top-squark pair production
$pp\to \tst_1\tst_1^*$\cite{Baer:2023uwo}.
Here, the HL-LHC reach extends to about $m_{\tst_1}\sim 1.9$ TeV whilst
naturalness allows for $m_{\tst_1}\sim 1-3$ TeV.
Gaugino (mainly wino) pair production at HL-LHC offers another promising
natural SUSY discovery channel\cite{Baer:2023olq}.

Beyond HL-LHC, CERN plans a future circular $e^+e^-$ collider (FCCee) in a
new 100 km tunnel which would operate as a Higgs (or $Z$) factory  with
$\sqrt{s}\sim 90-250$ GeV. While such a machine would have a higgsino
pair reach only to $m(higgsino)\sim 125$ GeV, it could also search for
light higgsinos via deviations in electroweak precision observables (EWPO)
like $\Delta m_W$ and $\Delta\sin^2\theta_{eff}$\cite{Nagata:2025ycf,Baer:2025tge}.
If such deviations in EWPO show up, they would point to new physics,
but the source of the new physics may not be clear. A higher energy
linear $e^+e^-$ collider with $\sqrt{s}>2m(higgsino)$ is also
a lucrative option for natural SUSY discovery\cite{ILC:2013jhg,Baer:2014yta,Baer:2019gvu,LinearColliderVision:2025hlt}.
The FCCee Higgs factory could be followed up at CERN by a new circular
$pp$ collider FCChh operating with $\sqrt{s}\sim 50-100$ TeV.

A more daring option is to pursue construction of a $\mu^+\mu^-$ collider.
This old idea\cite{Neuffer:1986dg,Barger:1996jm} has recently been pursued more strongly
by the US P5 report\cite{P5:2023wyd}
and the National Academy of Sciences (NAS) report\footnote{See
{\it National Academy of Sciences report: Elementary Particle Physics: The Higgs and Beyond}, https://nap.nationalacademies.org/resource/28839/EPP-one-pager.pdf.} on future
directions for high energy physics in the USA.
Most proponents envision a $\mu^+\mu^-$ collider operating with
center-of-mass (CoM) energy
in the $\sqrt{s}\sim 1-10$ TeV range\cite{AlAli:2021let}.
The technical challenge is to generate, cool and accelerate the $\mu^+$ and $\mu^-$ beams and bring them together
with a high enough luminosity, all before the beams decay.
The beam decay remnants could also pose possible health hazards that
would have to be overcome. The advantage of a muon collider of course is
that one can accelerate the muons to TeV-scale energies with little
energy loss from synchrotron radiation. Since the muons are fundamental,
all their annihilation energy can be transferred into new particle production.

Our goal in this paper is to examine what a TeV-scale muon collider can
do for the BSM case of natural supersymmetry.
Previous studies of $\mu^+\mu^-\to SUSY$ have surprisingly been absent or
else have examined rather implausible scenarios such as gluino pair
production\cite{AlAli:2021let}.
In the more plausible natural SUSY scenarios, a
plethora of reactions may be available to machines operating around
$\sqrt{s}\sim 10$ TeV. In such a case, it may pay to have flexible beam
energies which can allow the machine to scan over the various pair
production thresholds which are expected to turn on.
In addition, flexible beam energy could allow one to scan over heavy
Higgs resonances, so the machine becomes a heavy Higgs factory as envisioned
in Ref. \cite{Eichten:2013ckl}.
To this end, we introduce a natural SUSY (natSUSY) benchmark point
in Sec. \ref{sec:BM}.
The natSUSY BM point has sparticle mass spectra spread over 2 decades of
energy, with sparticle masses ranging from $0.2-30$ TeV.
In Sec. \ref{sec:natsusy}, we examine sparticle and Higgs pair production
and resonance production processes at a $\mu^+\mu^-$ collider.
We first focus on electroweakino pair production in Subsec. \ref{ssec:inos},
while in Subsec. \ref{ssec:squarks} we examine squark pair production and
in Subsec. \ref{ssec:higgs}  we examine Higgs pair production and
resonance production reactions. We conclude in Sec. \ref{sec:conclude}. 

\section{A plausible SUSY benchmark model}
\label{sec:BM}

Using the naturalness measure $\Delta_{EW}$, then it is found that many
formerly popular SUSY models are now ruled out by naturalness when the
light Higgs mass is restricted to be $m_h=125.5\pm 2.5$ GeV\cite{Baer:2014ica}.
This is the case for the CMSSM model\cite{Kane:1993td}, the mGMSB model\cite{Dine:1995ag}, mAMSB\cite{Randall:1998uk} and mirage mediation\cite{Choi:2005ge} (MM) for discrete choices of modular weights.
For the cases of mGMSB and mAMSB, the small $A_0$ values make it difficult to
accommodate $m_h\sim 125$ GeV unless 3rd generation matter scalars lie in the
tens-of-TeV regime. This leads to large $\Sigma_u^u(\tst_{1,2})$ contributions
to the weak scale and consequently large electroweak finetuning.
For the CMSSM model, low
$\mu\sim 100-300$ GeV is only found in the focus point
region\cite{Feng:1999zg}, where also low $A_0$ is required.
This region also has trouble generating $m_h\sim 125$ GeV
unless 3rd generation scalar masses become huge.

Fortunately, a model which allows for low $\Delta_{EW}$ and $m_h\sim 125$ GeV
is the original incarnation of the simplest possibility: gravity mediation
(or supergravity, SUGRA). Under SUGRA, scalar mass {\it non-universality}
is expected, unless some protective symmetry is acting, unlike the ad-hoc
universality which is input to the CMSSM.
Thus, in SUGRA, we expect intra-generational soft term universality
$m_{Q_i}=m_{U_i}=m_{D_i}=m_{L_i}=m_{E_i}\equiv m_0(i)$ since all elements of
each generation fill out the 16-dimensional spinor rep of $SO(10)$.
(Here we have in mind {\it local} GUTs where the GUT symmetry depends on
the geography of field locations on the compact manifold\cite{Nilles:2009yd}.)
This intra-generational universality also acts to reduce the finetuning contributions from a complete generation in the MSSM\cite{Baer:2013jla}.
Thus, in SUGRA we expect $m_{H_u}\ne m_{H_d}\ne m_0(1)\ne m_0(2)\ne m_0(3)$
but where $m_{H_u}\sim (1.3-2) m_0(3)$ is typically required for $m_{H_u}^2$
to run to natural, barely negative values\cite{Baer:2023cvi}.
Furthermore, in the context of the string landscape, as mentioned previously
there is a pull to large soft terms, but the pull on $m_0(1)$ and $m_0(2)$
has a generation independent upper bound arising from two-loop RG
effects\cite{Arkani-Hamed:1997opn}.
This leads to a decoupling/quasi-degeneracy landscape
solution to the SUSY flavor and CP problems which are otherwise endemic to
SUGRA models\cite{Baer:2019zfl}.
Thus, we expect $m_0(1)\sim m_0(2)\sim 10-50$ TeV.
In SUGRA, one also expects gaugino mass universality if each gaugino has
the same functional dependence on hidden sector fields
in the gauge kinetic function and also one expects
generically large $A_0\sim m_{3/2}$ soft terms which uplift $m_h\to 125$ GeV
via mixing effects and also lowers finetuning via cancellations in the
$\Sigma_u^u(\tst_{1,2} )$ terms\cite{Baer:2012up}.

These sorts of considerations can be realized in the 4-extra parameter
non-universal Higgs model NUHM4 with parameter choices
\be
m_0(i),\ m_{H_u},\ m_{H_d},\ m_{1/2},\ A_0,\ \tan\beta\ \ \ (NUHM4)
\ee
but where we will take $m_0(1)=m_0(2)=30$ TeV for simplicity, and since these
masses essentially decouple from our muon collider search considerations.
In NUHMi ($i=2-4$) models, it is convenient to trade $m_{H_u}$ and $m_{H_d}$
for the weak scale values $m_A$ and $\mu$ via the scalar potential
minimization conditions\cite{Ellis:2002wv}. Thus, we will work with NUHM3 with parameter set
\be
m_0(1,2),\ m_0(3),\ m_{1/2},\ A_0,\ \tan\beta,\ m_A,\ \mu\ \ \ (NUHM3) .
\ee
In Table \ref{tab:bm}, we adopt the NUHM3 parameter set from Ref. \cite{Baer:2024hpl}
with $m_0(1,2)=30$ TeV, $m_0(3)=6$ TeV, $m_{1/2}=2.2$ TeV, $A_0=-6$ TeV,
$\tan\beta =10$ with $\mu =200$ GeV and $m_A=2$ TeV.
The spectra are generated using Isajet 7.91\cite{Paige:2003mg} where
the value of $m_h$ typically agrees to be within $\pm 2$ GeV with corresponding
FeynHiggs\cite{Heinemeyer:1998yj} calculations\cite{Baer:2024fgd}.

From Table \ref{tab:bm}, we see the gluino mass $m_{\tg}\sim 5.2$ TeV
while first/second generation squarks and sleptons $m_{\tq ,\tell}\sim 30$
TeV so the model is safe from LHC gluino/squark search limits.
The lighter top-squark however is only $m_{\tst_1}\sim 1 $ TeV, at the edge of
present LHC exclusion limits using simplified models\cite{Canepa:2019hph,ATLAS:2024lda}.
The several higgsinos lie around $\sim 200$ GeV while heavy Higgs
$m_{H,A,H^\pm}\sim 2$ TeV. Thus, the predicted spectra is spread over two
orders of magnitude. In spite of this, the value of $\Delta_{EW}=25$
so the model is EW natural. With such heavy second generation sleptons, one
expects $(g-2)_\mu$ to be in accord with SM values\cite{Baer:2021aax}, which is borne out by
recent theory calculations\cite{Aliberti:2025beg}.

Invoking naturalness in the QCD sector, we expect the presence of a PQ sector
which would naturally be of DFSZ-form in SUSY (each requires two Higgs doublets).
Thus, one might expect mixed SUSY DFSZ axion plus higgsino-like WIMP
dark matter\cite{Bae:2013hma} where the relic abundance is typically
mainly axions\cite{Bae:2014rfa}.
Recent work solving the SUSY $\mu$ problem by first suppressing $\mu$ via
discrete $R$-symmetries\cite{Lee:2011dya}
(which naturally emerge from string compactifications\cite{Nilles:2017heg})
and then regenerating via $H_uH_d$ coupling to PQ sector fields
(Kim-Nilles\cite{Kim:1983dt}) can lead to small amounts of $R$-parity
violation so that the WIMPs that are produced in the early universe may all decay away
before the onset of Big Bang nucleosynthesis,
leaving axion-only dark matter\cite{Baer:2025oid,Baer:2025srs}
(seemingly in accord with recent strong limits of relic WIMP dark matter
from LZ\cite{LZ:2024zvo}). Thus, we expect our spectra to be safe from recent LHC searches and from recent WIMP direct detection searches.
%
\begin{table}\centering
\begin{tabular}{lc}
\hline
parameter & NUHM3 \\
\hline
$m_0(1,2)$   & 30000 \\
$m_0(3)$     & 6000  \\
$m_{1/2}$     & 2200 \\
$\tan\beta$  & 10 \\
$A_0$        & -6000 \\
\hline
$\mu$       & 200  \\
$m_A$       & 2000 \\
\hline
$m_{\tg}$    & 5178.7  \\
$m_{\tu_L}$  & 30256.0 \\
$m_{\tu_R}$  & 30323.3  \\
$m_{\te_R}$  & 29961.9  \\
$m_{\tst_1}$ & 1012.0  \\
$m_{\tst_2}$ & 3405.3  \\
$m_{\tb_1}$ & 3441.5 \\
$m_{\tb_2}$ & 5336.2  \\
$m_{\ttau_1}$ & 5639.4  \\
$m_{\ttau_2}$ & 5772.0  \\
$m_{\tnu_{\tau}}$ & 5725.2 \\
$m_{\tchi_2^\pm}$ & 1936.1  \\
$m_{\tchi_1^\pm}$ & 211.3  \\
$m_{\tchi_4^0}$  & 1938.8  \\ 
$m_{\tchi_3^0}$  & 1018.7  \\ 
$m_{\tchi_2^0}$  & 207.1  \\ 
$m_{\tchi_1^0}$  & 202.2  \\ 
$m_h$         & 124.5 \\ 
\hline
$\Omega_{\tchi_1^0}^{TP}h^2$ & 0.01 \\
$BF(b\to s\gamma)\times 10^4$  & 3.0  \\
$BF(B_s\to \mu^+\mu^-)\times 10^9$  & 3.8 \\
$\sigma^{SI}(\tchi_1^0, p)$ (pb) & $4.0\times 10^{-10}$ \\
$\sigma^{SD}(\tchi_1^0 p)$ (pb)  & $7.5\times 10^{-6}$  \\
$\langle\sigma v\rangle |_{v\to 0}$  (cm$^3$/sec)  & $2.0\times 10^{-25}$ \\
$\Delta_{\rm EW}$ & 25 \\
\hline
\end{tabular}
\caption{Input parameters and masses in~GeV units
  for an NUHM3 model   SUSY benchmark point
with $m_t=173.2$ GeV and generated from Isajet 7.91.
}
\label{tab:bm}
\end{table}

\section{Natural SUSY at a muon collider}
\label{sec:natsusy}

Next, we investigate what can be accomplished by a TeV-scale muon collider in the case of natural SUSY.

\subsection{Chargino and neutralino pair production}
\label{ssec:inos}

\begin{figure}[htb!]
\centering
    {\includegraphics[height=0.4\textheight]{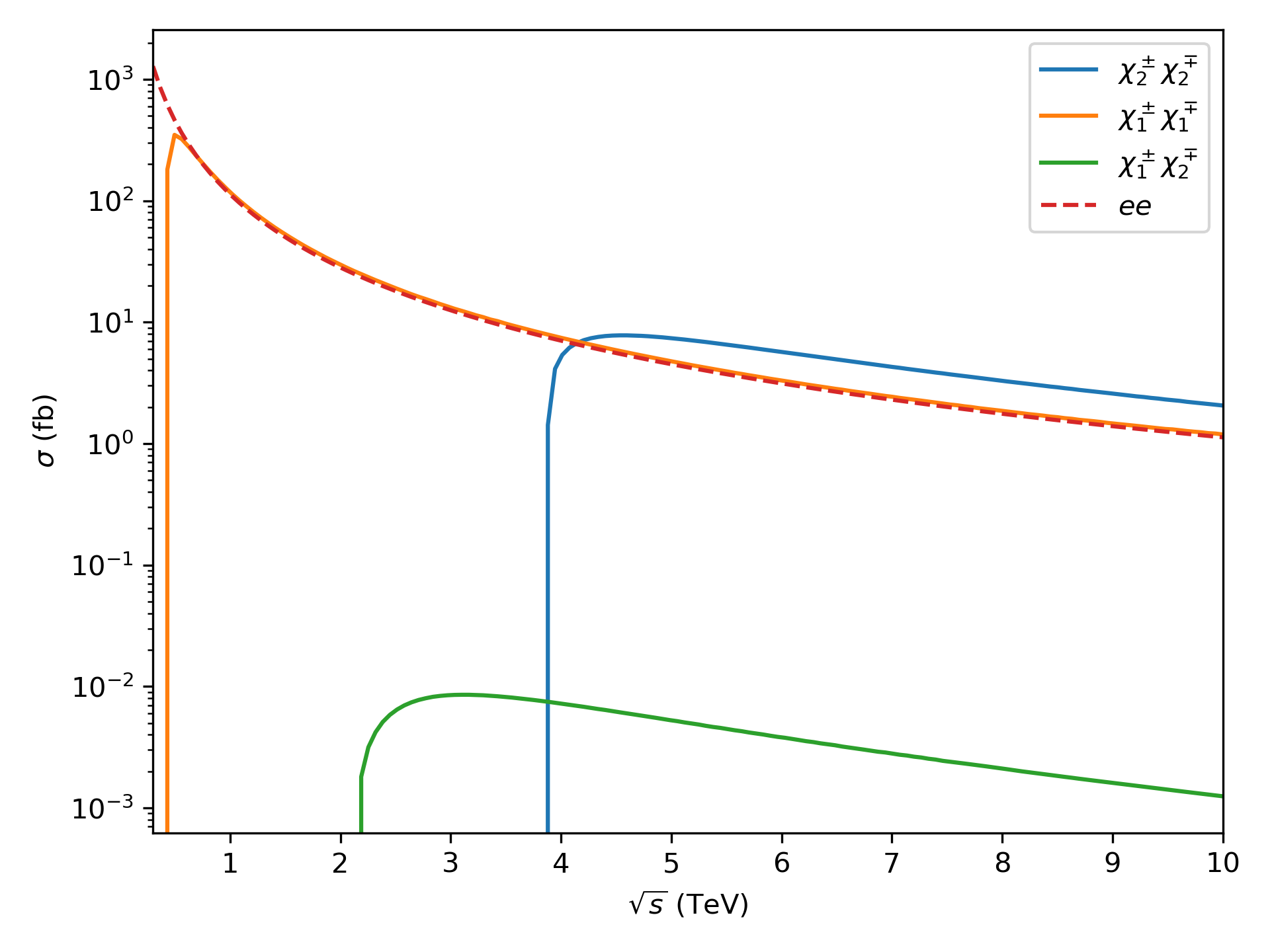}}\\
        {\includegraphics[height=0.4\textheight]{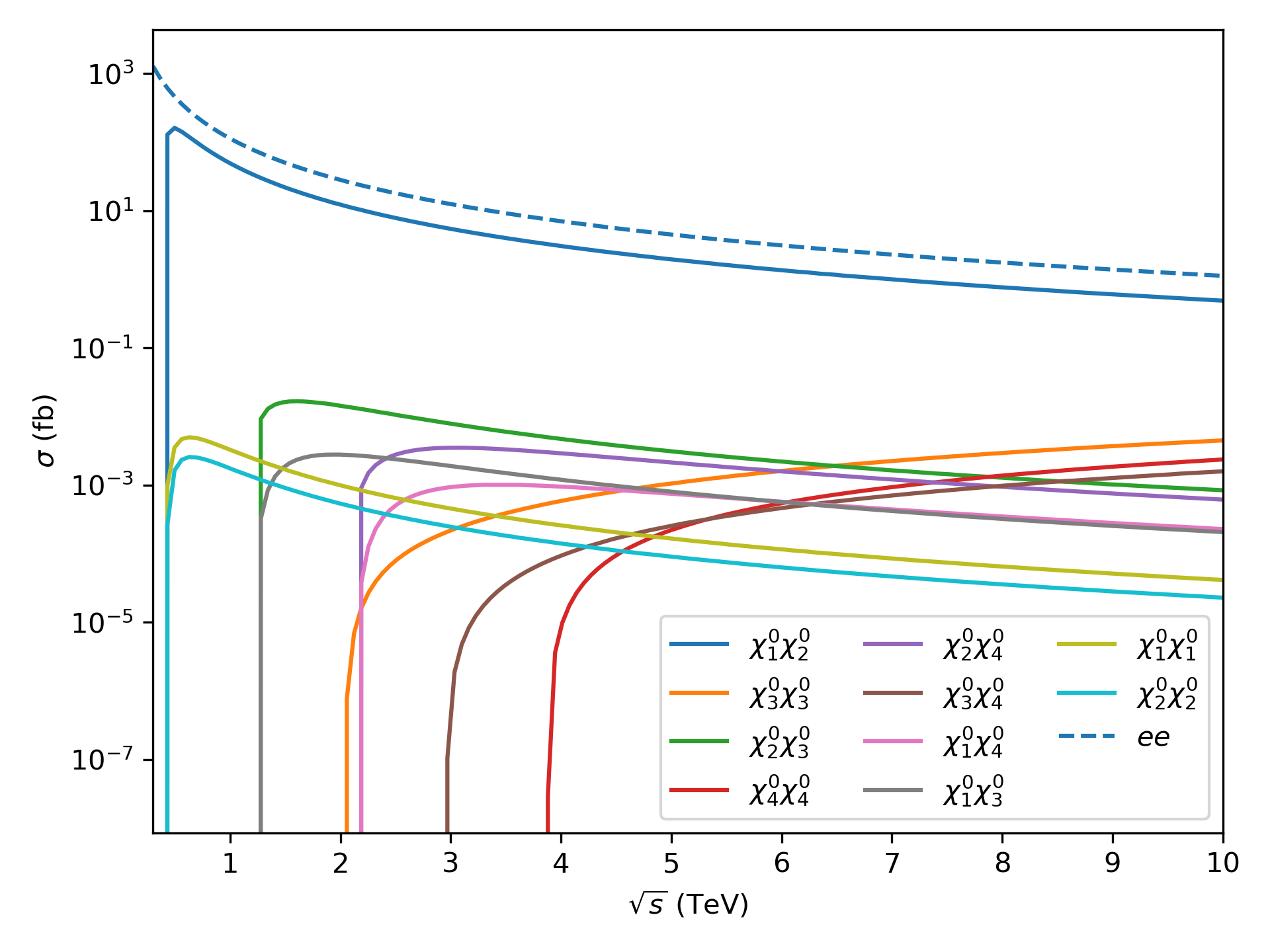}}
        \caption{Plot of {\it a}) chargino pair production rates vs. $\sqrt{s}$
          and {\it b}) neutralino pair production vs. $\sqrt{s}$
          at a $\mu^+\mu^-$ collider for the natural SUSY benchmark point
          listed in the text.
      \label{fig:inos}}
\end{figure}

We first show total cross section rates $\mu^+\mu^-\to SUSY$ for the case of
electroweakinos, since these include the vaunted light higgsinos
which are the hallmark of natSUSY. In Fig. \ref{fig:inos}{\it a}),
we show the chargino pair production rates
$\sigma (\mu^+\mu^-\to\tchi_1^+\tchi_1^-,\ \tchi_1^\pm\tchi_2^\mp$ and
$\tchi_2^+\tchi_2^-$). As a reference cross section, we also show
the rate for $\mu^+\mu^-\to e^+e^-$ (dashed red curve). The threshold
for charged higgsino pair production begins around
$\sqrt{s}=2m(higgsino)\sim 400$ GeV where $\tchi_1^+\tchi_1^-$ production
followed by $\tchi_1^-\to f\bar{f}^\prime\tchi_1^0$ decay (where $f$
and $f^\prime$ are SM fermions).
The $\tchi_1^\pm$ decay products are very soft near threshold since
most of the decaying particle rest energy goes into making $m_{\tchi_1^0}$,
and the $\tchi_1^0$ LSP is expected to yield missing energy,
even in the case of late-decaying $\tchi_1^0$.
The $\tchi_1^+\tchi_1^-$ cross section is nearly the same as for $e^+e^-$
production since both processes are dominated by $s$-channel $\gamma$ exchange.
As the beam energy increases, the decay products will become more energetic
and collimated due to the increased production energy of the $\tchi_1^\pm$s.

As $\sqrt{s}$ increases, one passes the $\mu^+\mu^-\to\tchi_1^\pm\tchi_2^\mp$
threshold at just over $\sqrt{s}\sim 2$ TeV where charged higgsino+wino
production begins. This reaction takes place just via $s$ channel $Z$
exchange and so is highly suppressed by about three orders of magnitude
compared to $\tchi_1^+\tchi_1^-$ production; it may not even be visible
depending on the integrated luminosity which can be achieved for a muon
collider.
In Ref. \cite{AlAli:2021let}, integrated luminosity values of $0.2-10$ ab$^{-1}$
are entertained for $\sqrt{s}$ varying from $1-10$ TeV.
Thus, we might expect just of order $1-10$ events of the mixed
$\tchi_1^\pm\tchi_2^\mp$ chargino pair production reactions.
Near $\sqrt{s}\sim 4$ TeV, then $\tchi_2^+\tchi_2^-$
(charged wino pair production) turns on.
These reaction rates can exceed even the benchmark $\mu^+\mu^-\to e^+e^-$
process. The charged winos decay via $\tchi_2^-\to W^-\tchi_{1,2}^0$,
$\tchi_1^- Z^0$ and $\tchi_1^- h$ each at around 25\%.
The $WW$, $WZ$, $ZZ$, $Wh$ and $Zh+\eslt$ final states should be
easily recognizable. If variable beam energy is possible, then threshold scans
could determine $m(\tchi_2^\pm)$ to good precision.

In Fig. \ref{fig:inos}{\it b}), we show the rates for neutralino
pair production $\tchi_i^0\tchi_j^0$ with $i,j=1-4$ (ten reactions total).
The reactions again turn on around $\sqrt{s}\sim 2\mu$ where
neutral higgsino pair production begins.
Only the reaction $\tchi_1^0\tchi_2^0$ exceeds the fb level since
production is only via $s$-channel $Z$ exchange.
It is comparable to but less than the benchmark $e^+e^-$
production rate.
The virtual $Z$ only couples to the higgsino components of the
neutralinos so the neutral gaugino pair reactions are suppressed while
cancellations in the $Z\tchi_i^0\tchi_i^0$ couplings suppress
the $\tchi_1^0\tchi_1^0$ and $\tchi_2^0\tchi_2^0$ production
rates\cite{Baer:2006rs}.

The $\tchi_2^0\to f\bar{f}\tchi_1^0$ so the invariant mass $m(f\bar{f})$
is kinematically bounded by $m_{\tchi_2^0}-m_{\tchi_1^0}$ which presents a
distinctive signature.
As the CoM energy $\sqrt{s}$ increases beyond threshold,
the energy of the $f\bar{f}$ final state increases whilst its invariant
mass remains kinematically fixed. A variety of other $\tchi_i^0\tchi_j^0$
reactions turn on starting at $\sqrt{s}\sim 2$ TeV where neutralino
higgsino+wino begins. These all present interesting signatures
since $\tchi_3^0\to \tchi_1^\pm W^\mp$ at $\sim 25\%$ each, along with
$\tchi_2^0 Z$ and $\tchi_1^0 h$, while $\tchi_4^0\to \tchi_1^\pm W^\mp$
at $\sim 25\%$, and $\tchi_2^0 Z$ and $\tchi_2^0 h$ at $\sim 20\%$.
However, the individual reactions may be difficult to pick out since
we expect only a handful of each to be produced at the projected
integrated luminosities.

\subsection{Squark pair production}
\label{ssec:squarks}

\begin{figure}[htb!]
\centering
    {\includegraphics[height=0.5\textheight]{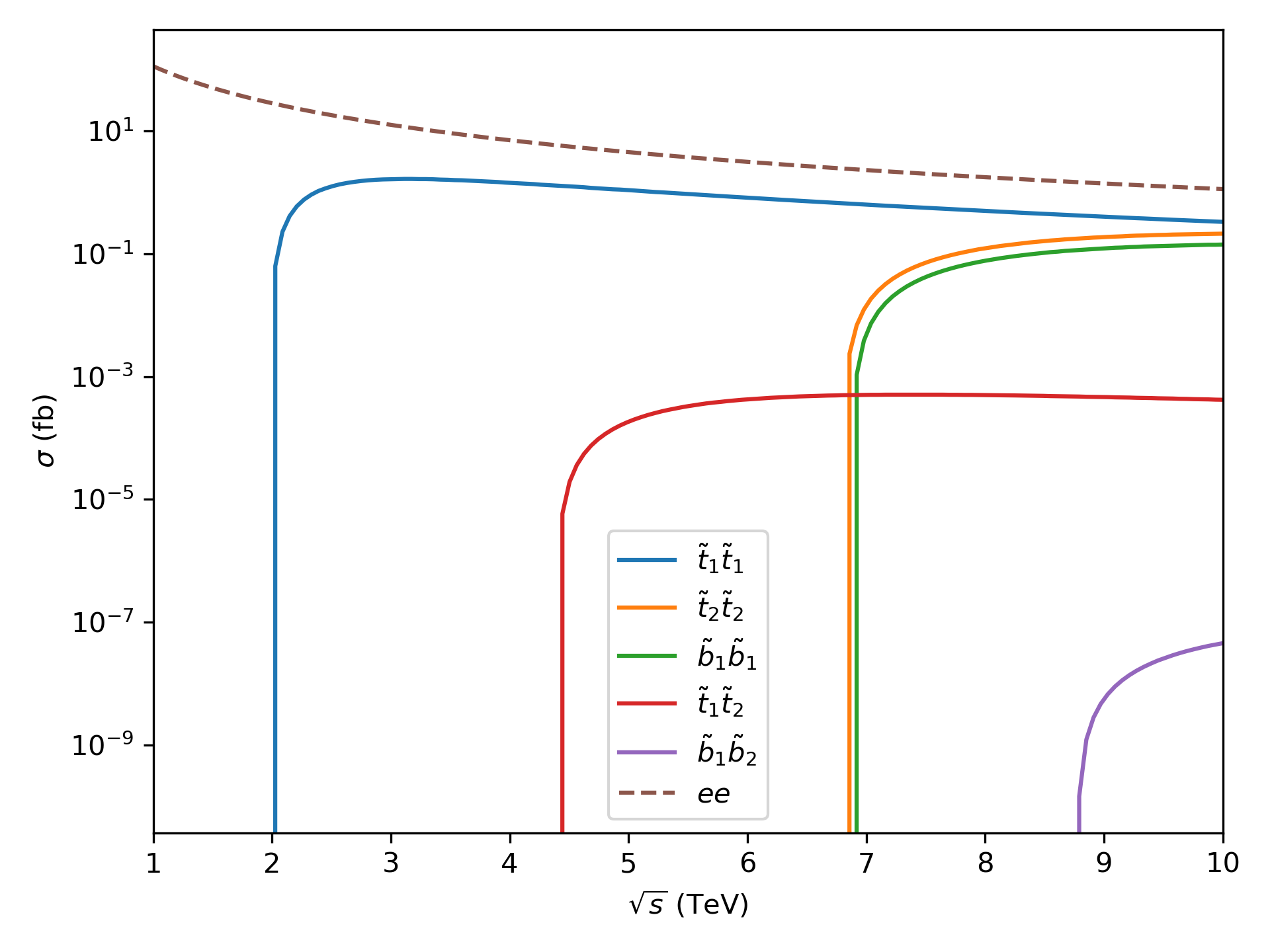}}
    \caption{Plot of {\it a}) stop/sbottom pair production
      cross sections vs. $\sqrt{s}$
          at a $\mu^+\mu^-$ collider for the natural SUSY benchmark point
          listed in the text.
      \label{fig:squarks}}
\end{figure}

While first/second generation sfermion pair production would not be accessible
even to the highest energy muon colliders, the third generation
squark pair production would be available. In Fig. \ref{fig:squarks},
we show the cross sections for $\tst_1\tst_1^*$, $\tst_2\tst_2^*$,
$\tst_1\tst_2^* +\tst_2\tst_1^*$, $\tb_1\tb_1^*$ and $\tb_1\tb_2^*+\tb_2\tb_1^*$
production vs. $\sqrt{s}$. Naturalness requires the $\tst_1$ to be the lightest
squarks for the NUHM3 model, and the $\mu^+\mu^-\to\tst_1\tst_1^*$
reaction threshold appears around $\sqrt{s}\sim 2$ TeV for our BM model.
In this case, $\tst_1\to b\tchi_1^+$ at $\sim 50\%$ followed by
$t\tchi_{1,2}^0$ at $\sim 25\%$ each. Thus, from light stop pair
production, we expect final states of $b\bar{b}$, $t\bar{t}$ and
$t\bar{b}/\bar{t} b$ plus light higgsino pairs, where the latter yield
softer decay products along with large $\eslt$. The $\tst_1\tst_1^*$ reactions
occur at nearly the rate for $e^+e^-$ production once production threshold
is passed. As beam energy increases, the $\tst_1\tst_2$ threshold is passed,
but these reactions only occur at the $\sim 10^{-3}$ fb level. Here, $\tst_2\to tZ$ and $th$ at $\sim 30\%$ and $\sim 40\%$ respectively.
Around $\sqrt{s}\sim 7$ TeV, both $\tst_2\tst_2^*$ and $\tb_1\tb_1^*$ turn on
and quickly approach the fb level cross sections.
The $\tb\to \tst_1 W$ and $\tchi_1^- t$ at 68\% and 21\%
respectively. The $\tb_1\tb_2^*+\tb_2\tb_1^*$ reactions turn on around
$\sqrt{s}\sim 9$ TeV but at likely too low a rate to be observed. However, owing to the clean environment of a muon collider, these hadronic states stand out more distinctly than at hadronic colliders such as the LHC or FCC-hh. As the beam energy increases, the decay products of $\tilde{t}$ and $\tilde{b}$ become increasingly collimated, allowing clearer identification of the individual squark origins.

\subsection{SUSY Higgs boson production}
\label{ssec:higgs}
Heavy Higgs boson production is particularly exciting at a $\mu^+\mu^-$
collider\cite{Gunion:1989we}.
From Eq. \ref{eq:mzs}, the heavy Higgs contributions to $\Delta_{EW}$
are $(m_{H_d}/\tan\beta)^2/(m_Z^2/2)$ so for $\tan\beta\sim 10$ and
$\Delta_{EW}\alt 30$ and with $m_{A/H}\sim m_{H_d}$ we expect
$m_{H/A/H^\pm}\alt 3.5$ TeV\cite{Bae:2014fsa}.

\begin{figure}[htb!]
\centering
        {\includegraphics[height=0.5\textheight]{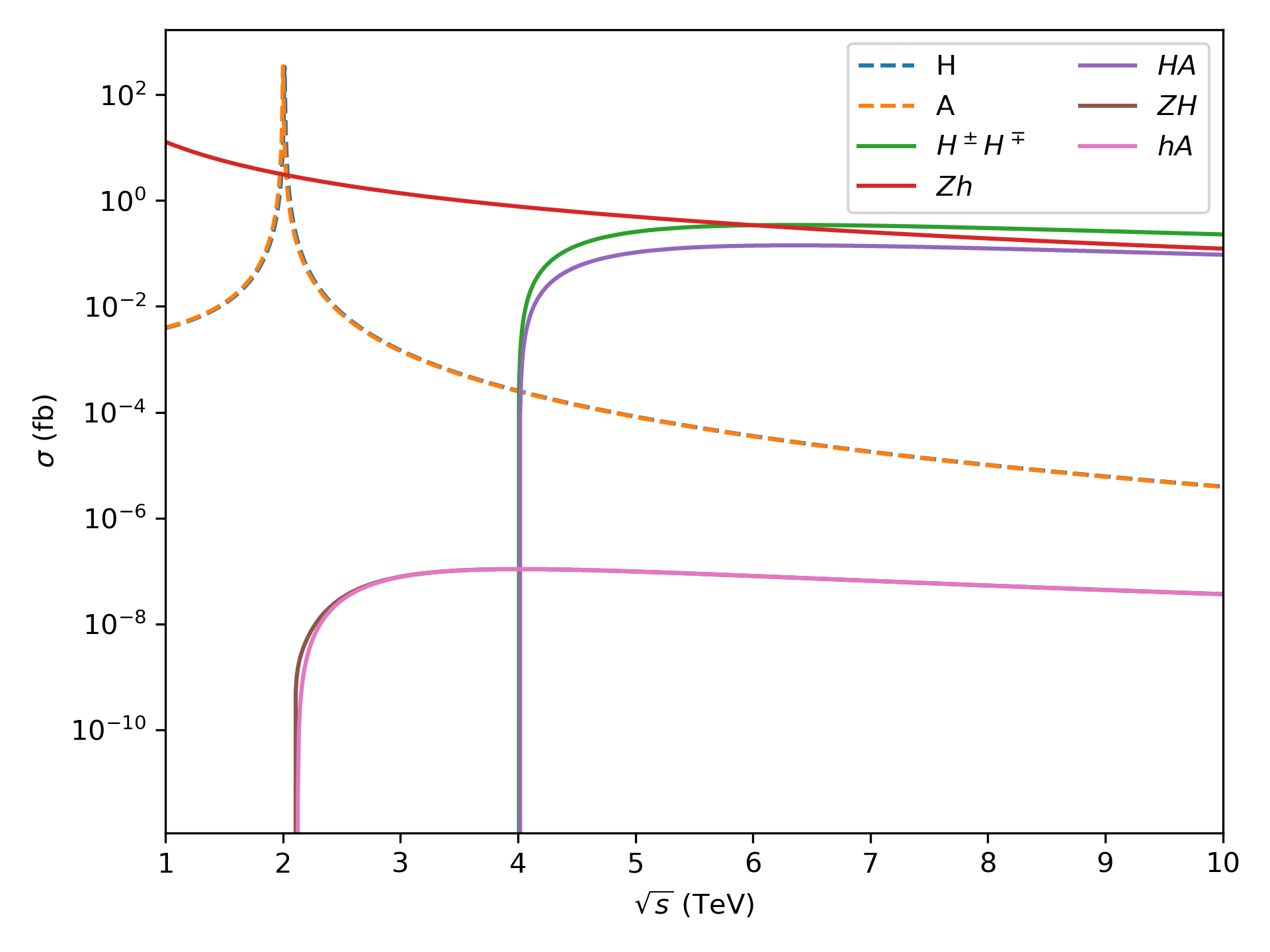}}
        \caption{Plot of Higgs boson production cross sections vs. $\sqrt{s}$
          at a $\mu^+\mu^-$ collider for the natural SUSY benchmark point
          listed in the text.
      \label{fig:higgs}}
\end{figure}

The $s$-channel resonance production of heavy neutral Higgs bosons is given by
\be
\sigma (\mu^+\mu^- \to H/A )=\frac{4\pi\Gamma_{H/A}^2 BF(H/A\to\mu^+\mu^-)}{(s-m_{H/A}^2)^2+\Gamma_{H/A}^2m_{H/A}^2}
\ee
and is plotted in Fig. \ref{fig:higgs} where the $H$ and $A$ resonances
are clearly visible near $\sqrt{s}\sim 2$ TeV.
The $H/A$ cross sections at resonance exceed the baseline $\mu^+\mu^-\to Zh$
value. Having variable muon beam energy in the vicinity of $H/A$
resonance production would allow good precision in mapping out the
heavy neutral Higgs properties. And if sufficient precision on beam energy
is attained, one might even separate out the two (overlapping) $H$ and $A$
resonances\cite{Eichten:2013ckl}.

Shortly beyond $H/A$ resonance production, the $\mu^+\mu^-\to hA$ and $ZH$
reactions turn on, albeit at likely unobservable levels.
At $\sqrt{s}\agt 4$ TeV, then the $\mu^+\mu^-\to HA$ and $H^+H^-$
reactions turn on. The charged Higgs pair production reaction is
quite difficult to see at hadron colliders\cite{Baer:2023yxk}, but occurs at a
high rate at a TeV-scale $\mu^+\mu^-$ collider.
In natural SUSY, since one expects $m_{H/A/H^\pm}\gg 2\mu$, then SUSY
decay modes of the heavy Higgs states are always open.
These decay modes are maximal for mixed higgsino/gaugino decays and
can dominate\cite{Baer:2022smj} the SM decay modes
(which are usually assumed in most heavy Higgs analyses).
For instance, for our BM scenario, the charged Higgs branching fraction
$H^+\to t\bar{b}$ is actually reduced to just 61\% while the decay
$H^+\to \tchi_1^+\tchi_3^0$ is lifted to 29\%.

\section{Conclusions}
\label{sec:conclude}

Natural supersymmetry with low $\Delta_{EW}$ is a highly plausible extension
of the SM which 1. solves the gauge hierarchy problem 2. explains why the
weak scale $m_{W,Z,h}\sim 100$ GeV (because all contributions to the weak scale are comparable to or smaller than this value) and 3. is supported by an array of
virtual quantum effects.
NatSUSY in a landscape context supports and even predicts that $m_h\sim 125$ GeV with sparticles beyond present LHC limits.
Previous work has examined in great detail how natSUSY might manifest itself at HL-LHC and linear $e^+e^-$ colliders, but little or no work has been done for muon colliders, even though work towards realizing a muon collider has become a high priority item on future HEP facilities wish list.

Here we remedy this situation by examining what a muon collider with
$\sqrt{s}=1-10$ TeV can do in the natSUSY context.
Using a very plausible natSUSY benchmark
point which includes a solution to the SUSY flavor/CP problems, we show that a
variety of SUSY pair production reactions should be available to a muon collider, starting with higgsino pair production for $\sqrt{s}\agt 2|\mu |$.
Mixed higgsino/gaugino and gaugino pair production will allow access to the heavier electroweakino pair production reactions. Given control over the muon collider beam energy, then precision scans at or around threshold production may allow for precision sparticle mass measurements which are otherwise rather difficult at hadron colliders due to the presence of $\eslt$. Third generation squark pair production should also be accessible since top squark masses are bounded in natSUSY at the several TeV level. The capability of a $\mu^+\mu^-$ collider
really shines for heavy Higgs production, where in natSUSY we expect heavy Higgs states with mass $\alt$ several TeV. With variable beam energy, the
resonant production of the $H/A$ states may be explored. At higher energies,
$HA$ and $H^+H^-$ production should open up.
Mapping out the threshold for charged Higgs pair production at a muon collider
should allow precision measurement of $m_{H^\pm}$ which is otherwise extremely
difficult at hadron collider (and likely beyond reach of linear $e^+e^-$
colliders).

{\it Acknowledgements:} 

VB gratefully acknowledges support from the William F. Vilas estate.
HB gratefully acknowledges support from the Avenir Foundation.


\bibliography{muonc}
\bibliographystyle{elsarticle-num}

\end{document}